\documentclass[10pt, conference, letterpaper]{IEEEtran}
\IEEEoverridecommandlockouts
\usepackage{cite}
\usepackage{amsmath,amssymb,amsfonts}
\usepackage{algorithmic}
\usepackage{graphicx}
\usepackage{textcomp}
\usepackage{xcolor}
\usepackage{xurl}
\usepackage{setspace}
\usepackage{subfiles}
\usepackage{soul} 
\usepackage{color, xcolor} 
\soulregister{\cite}7 
\soulregister{\eqref}7 
\usepackage{algorithm}
\usepackage{amsthm}
\newtheorem{theorem}{Theorem}

\newtheorem{lemma}{Lemma}

\allowdisplaybreaks[2]
\def\BibTeX{{\rm B\kern-.05em{\sc i\kern-.025em b}\kern-.08em
    T\kern-.1667em\lower.7ex\hbox{E}\kern-.125emX}}

\usepackage{graphicx} 
\usepackage{subfigure}
\usepackage{subcaption} 
\begin{document}
\pagestyle{plain}
\title{The Space above the Sky: Uniting Global-Scale Ground Station as a Service for 
Efficient Orbital Data Processing\\

\thanks{This work is supported by the National Natural Science Foundation of
China (Grant No. 62302292) and the Fundamental Research Funds for the
Central Universities. Yifei Zhu is the 
corresponding author.}
}

\author{\IEEEauthorblockN{Heng Zhao\IEEEauthorrefmark{1}, Sheng Cen\IEEEauthorrefmark{1}, Yifei Zhu\IEEEauthorrefmark{1}\IEEEauthorrefmark{2}}
\IEEEauthorblockA{\IEEEauthorrefmark{1} UM-SJTU Joint Institute, Shanghai Jiao Tong University, China}
\IEEEauthorblockA{\IEEEauthorrefmark{2}Cooperative Medianet Innovation Center (CMIC), Shanghai Jiao Tong University, China}
\IEEEauthorblockA{
Email: zhao-heng@sjtu.edu.cn, cens98@sjtu.edu.cn, yifei.zhu@sjtu.edu.cn}
}

\maketitle
 
\begin{abstract}

Large constellations of Earth Observation Low Earth Orbit satellites collect enormous amounts of image data every day. This amount of data needs to be transferred to data centers for processing via ground stations. Ground Station as a Service (GSaaS) emerges as a new cloud service to offer satellite operators easy access to a network of ground stations on a pay-per-use basis. However, renting ground station and data center resources still incurs considerable costs, especially for large satellite constellations. The current practice of sticking to a single GSaaS provider also suffers high data latency and low robustness to weather variability due to limited ground station availability. To address these limitations, we propose SkyGS, a system that schedules both communication and computation by federating GSaaS and cloud computing services across multiple cloud providers. We formulate the resulting problem as a system cost minimization problem with a long-term data latency threshold constraint. In SkyGS, we apply Lyapunov optimization to decompose the long-term optimization problem into a series of real-time optimization problems that do not require prior knowledge. As the decomposed problem is still of exponential complexity, we transform it into a bipartite graph-matching problem and employ the Hungarian algorithm to solve it. We analyze the performance theoretically and evaluate SkyGS using realistic simulations based on real-world satellites, ground stations, and data centers data. The comprehensive experiments demonstrate that SkyGS can achieve cost savings by up to 63\% \& reduce average data latency by up to 95\%.
 
\end{abstract}


\section{Introduction}
Due to the advancements in satellite technology and declining costs of accessing space, Low Earth Orbit (LEO) satellites have witnessed a significant upsurge over the past two decades, reaching almost seven thousand as of 2023\cite{UCS_Satellite_Database}. Among these satellites, about 45\% are Earth observation satellites that operate at altitudes under 1,000 kilometers and capture detailed imaging data of the Earth’s activities. Many companies including Spire Global\cite{spire} and Planet\cite{planet} have launched constellations containing hundreds of these satellites, enabling consistent, high-resolution surveillance of the Earth. Earth observation LEO satellites are critical tools for earth monitoring and have been widely applied in agriculture\cite{aragon2018cubesats, nguyen2020monitoring}, disaster monitoring\cite{lei2022flood, chuvieco2020satellite}, and natural resource management\cite{sheffield2018satellite, mahdavi2018remote}.
 
Each satellite from these constellations collects around one terabyte of data every day\cite{tao2023transmitting}. The immense volume of data is transmitted back to Earth via ground stations and sent to the data centers for advanced processing. Making such communication possible requires significant financial investment to construct a ground station, including costs for purchasing land, buildings, and hardware\cite{ground-station-cost}. To mitigate these costs and lower the barriers to entry for accessing satellites, ground station providers, including some cloud providers, like AWS \cite{AWS}, SSC \cite{ssc_ground_station}, and Leaf Space \cite{leafspaceservices} offer Ground Station as a Service (GSaaS). This service allows satellite operators to utilize an existing network of ground stations without the need for direct ownership or operational management of the infrastructure. For example, Astrocast utilizes Leaf Space's GSaaS to support its Global IoT constellation without building its ground station networks \cite{leaf_space_Astrocast}. 
 
The transient nature of satellite communications between LEO satellites and ground stations necessitates robust, densely arranged, and geographically distributed ground stations for low-latency data transmission. However, many satellite operators today only depend on a single ground station provider, which often fails to offer an adequate number of ground stations necessary for low-latency communication. In particular, compared with large satellite constellations, which often include more than 100 satellites, the number of ground stations provided by a single provider is relatively small. For instance, Leaf Space operates 17 ground stations \cite{leafline}, while SSC operates 10 ground stations \cite{ssc_ground_station}. In contrast, the Dove constellation, managed by Planet, comprises over 150 Earth observation LEO satellites\cite{planet}. This disparity leads to significant satellite contention and consequently, high data latency. Often, multiple satellites are within the visibility range of one ground station simultaneously, competing for the limited downlink opportunities. Given the scarcity of these opportunities, unmanaged contention can result in substantial data retention onboard the satellites. 
 
This latency is particularly critical for time-sensitive applications such as flood monitoring \cite{lei2022flood} and forest fire detection \cite{chuvieco2020satellite}. Moreover, relying on a single ground station provider decreases the robustness of satellite communications against weather variations. The quality of the links between satellites and ground stations is highly susceptible to atmospheric conditions. Attenuation effects, which can be influenced by rain and clouds, may result in signal losses of up to 10 dB, varying with the weather conditions and the signal's frequency \cite{shrestha2017characterization}. Without additional ground stations serving as backups, satellites must contend with these weather-induced attenuation during data downlink, potentially compromising the reliability and quality of the communication.

Following the idea of sky computing\cite{stoica2021cloud}, it is interesting to extend the benefit of sky computing to outer space and examine the possibility of sky GSaaS. Through federating multiple GSaaS from diverse providers, the scale of ground station networks is significantly enhanced. Satellites can downlink data to any station within this collective network, thereby substantially increasing downlink opportunities and reducing data onboard waiting time.
 
Despite the huge potential benefits, realizing the federation of GSaaS is non-trivial and faces the following challenges: 
\begin{itemize}
    \item \textbf{Compounding interaction of communication and computing:} In satellite data downlink scheduling, current methods primarily focus on optimizing the communication aspect, often neglecting the data processing stage. For instance, L2D2 optimizes data downlink scheduling between satellites and ground stations\cite{vasisht2021l2d2}, while Umbra considers scheduling data from satellites to ground stations and then to the cloud, without addressing subsequent data processing \cite{tao2023transmitting}. To be comprehensive, incorporating satellite data processing into downlink scheduling is essential. However, this integration introduces significant complexity. Challenges include the transfer of data from ground stations to data centers, the rental of data centers and computational time required for processing, and the potential for faster transmission links between adjacent data centers and ground stations, all of which contribute to increased system complexity. 
    \item \textbf{Complex downlink scheduling:} Given the vast number of satellites, ground stations, and data centers, dynamically pairing them requires considering various factors such as satellite orbits, data backlog, ground-satellite link (GSL) quality, and computation times. When multiple satellites compete for a single ground station, selecting the most suitable satellite is crucial for optimizing system performance. This pairing process is inherently complex, especially when we consider the trade-off between system cost and data latency: frequent downlinks ensure low latency but increase costs while accumulating data onboard and fully utilizing the transmission link during passes minimizes costs but increases latency. Balancing this trade-off is a significant challenge in downlink scheduling.
\end{itemize}

In this paper, we introduce the SkyGS system, which comprehensively integrates communication and computation aspects of satellite data. The system employs a broker to effectively pair satellites, ground stations, and data centers. SkyGS federates multiple GSaaS from diverse providers and data centers across multiple clouds for data reception and processing. Key benefits of the SkyGS architecture include a denser ground station network, increased downlink opportunities for satellite data, reduced onboard awaiting time, and enhanced freedom in data center selection, which facilitates faster and more cost-effective data processing.
 
SkyGS adopts the Lyapunov optimization framework to determine the pairing of satellites, ground stations, and data centers. Our approach focuses on minimizing system costs over the long run while maintaining a predefined long-term data latency threshold. By applying the Lyapunov optimization technique, SkyGS can effectively incorporate the long-term latency threshold into real-time optimizations and make online decisions on dynamic pairing without requiring any future information (e.g., satellite passes and GSL quality). To address the challenge of exponential complexity of the resulting real-time optimizations, we transform the problem into a bipartite graph and apply the Hungarian algorithm to solve it, thereby reducing the computational complexity to $O(|E||\mathcal{D}| + (|\mathcal{S}|+|\mathcal{A}|)^3)$, where $|E|$, $|\mathcal{D}|$, $|\mathcal{S}|$ and $|\mathcal{A}|$ denote the number of edges, data centers, satellites, and antennas, respectively. In summary, our main contributions are summarized as follows:
\begin{itemize}
    \item We propose a novel concept to federate multiple GSaaS from diverse providers into a unified ground station network.
    \item During downlink scheduling, we broaden our focus from solely communication to encompass both communication and computation, thereby incorporating satellite data processing into our considerations.
    \item During optimization, SkyGS judiciously considers both the system cost and data latency and effectively minimizes system cost while adhering to the data latency threshold.
    \item Our methodology is based on the Lyapunov optimization framework, which ensures the attainment of a solution that closely approximates the optimal, within a verifiable upper boundary. 
    \item We propose transforming the initial online optimization problem, which has exponential complexity, into a bipartite graph matching framework. By employing the Hungarian algorithm, we substantially reduce the computational complexity to a polynomial level.
    \item After evaluation, our results demonstrate that SkyGS can achieve cost savings of up to 63\% and reduce average data latency by up to 95\%. 
\end{itemize}

\section{Background}
\noindent\textbf{Satellite orbits and data collection:}
Emerging LEO constellations for earth observation, such as the Dove constellation, typically operate in polar orbits with an approximate period of 90 minutes. Due to the Earth's rotation, the satellite's ground track shifts westward with each orbit, enabling the scanning of different global sections each time\cite{EO_wiki}. Equipped with sensors, these satellites capture earth images across various parts of the frequency spectrum, including RGB, radio waves, and infrared. Unlike traditional Earth observation satellites that prioritize and plan their imagery collections based on specific "targeted" areas, often omitting non-prioritized regions\cite{Dove-PDF-Download}, emerging Earth observation LEO constellations aim to build a near real-time map of the earth. These satellites continuously collect data, storing images on-board. As measured in work\cite{tao2023transmitting}, the Dove constellation captures an average of 120TB of data per day. Handling the downlink and processing of such immense volumes of data incurs substantial costs and requires timely execution, underscoring the necessity of considering both system cost and data latency during scheduling.

\noindent\textbf{Data downlink scheduling:}
Transmitting image data to data centers involves two stages: from the satellite to a ground station, and then from the ground station to the data center. 

During the initial stage, image data remain onboard the satellite until it establishes contact with a ground station. A single ground station equipped with multiple antennas can communicate with several satellites concurrently, with each antenna dedicated to one satellite. However, the decision to downlink data is influenced by several factors, even when satellites are within the ground station's range. Contention is a common issue. As shown in work\cite{vasisht2021l2d2}, a ground station can simultaneously detect up to 100 satellites. Even when satellites are spaced evenly in orbit, they may compete for the same ground stations\cite{foster2015orbit}. When multiple satellites vie for communication, priority is generally assigned based on several criteria including the amount of data backlog onboard and the quality of the GSL. Since renting a ground station antenna can cost 22 dollars per minute\cite{velusamy2022ai}, operators may choose to skip some satellite passes to reduce costs, allowing data to accumulate. While this strategy is cost-effective, it leads to increased data latency.

In the second stage of data transmission, image data is transferred from ground stations to data centers via backhaul links. Ground stations are typically situated in remote or rural locations to minimize potential electromagnetic interference and to provide unobstructed visibility of the sky\cite{alen_gs_position_selection}, which is crucial for effective satellite communication. While this positioning optimizes the conditions for satellite links, it presents significant challenges for the backhaul infrastructure. The remote nature of these stations often means that the bandwidth available for backhaul links is limited, which in turn prolongs the latency of data transfer from the ground stations to the data centers \cite{tao2023transmitting}.

\noindent\textbf{Ground Station as a Service(GSaaS):} 
For satellite operators, GSaaS emerges as a cost-effective alternative to the high expense of constructing ground station infrastructure. GSaaS represents a fully managed service, empowering operators to handle satellite communications without the burden of building or maintaining ground station infrastructure\cite{AWS}. Typically, GSaaS providers charge for ground station antenna rental by the minute, with diverse pricing options to accommodate different operational needs. Amazon Web Services (AWS), for instance, provides two pricing options: On-Demand, where you pay as you use with no long-term commitments, and Reserved, offering lower rates for a monthly or 12-month commitment, suitable for regular use\cite{AWS_price}. 

Considering there is no ingress charge for data transfers into a cloud, we neglect the transfer cost from the ground station to the data center. Satellite data is processed in data centers, where the prices of resources, such as virtual machines (VMs), are usually varied in both spatial and temporal domain\cite{xiao2017cost}. According to Amazon's real pricing model\cite{aws_ec2}, costs are calculated per instance-hour from launch to termination, with partial hours billed per second for certain VM types.

\section{Related Work}
\noindent\textbf{Satellite Downlink Scheduling:} Existing works\cite{tao2024known, vasisht2021l2d2, lai2021orbitcast, tao2023transmitting, augenstein2016optimal} on satellite downlink scheduling are dedicated to minimize data latency\cite{tao2024known, vasisht2021l2d2, lai2021orbitcast}, maximize throughput\cite{tao2023transmitting, augenstein2016optimal}. Vasisht et al.\cite{vasisht2021l2d2} propose a distributed and hybrid ground station model, and utilize the Hungarian algorithm to identify the optimal match between satellites and ground stations, thereby reducing downlink latency and enhancing communication robustness. Addressing the issue of throughput loss and high data latency due to uneven queuing effect, researchers in \cite{tao2023transmitting} present a withhold scheduling algorithm based on time-expanded networks. Additionally, Lai et al.\cite{lai2021orbitcast} propose a hybrid architecture for delivering Earth observation data, leveraging LEO constellations and geo-distributed ground stations to enable low-latency and scalable data transmission from space. While several studies make great contributions to data latency reduction and throughput improvement for image data downlink, they predominantly focused on communication challenges, overlooking the data processing process. Different from them, this paper aims to broaden the perspective by integrating computational considerations into downlink scheduling. Besides scheduling data from satellites to ground stations, we route data from ground stations to data centers across different clouds for data processing. 

\noindent\textbf{Space-air-ground integrated networks(SAGINs):}
In recent years, SAGINs have emerged as a significant technological advancement, integrating satellite, aerial, and terrestrial networks to offer comprehensive connectivity solutions\cite{liu2018space}. Recent works\cite{zhang2022space, kato2019optimizing, jia2021toward, chen2021civil, ye2020space} has investigated resource allocation\cite{zhang2022space},  data collection and transmission\cite{jia2021toward}, outage analysis\cite{ye2020space} and system optimization\cite{kato2019optimizing, chen2021civil}. Researchers in \cite{kato2019optimizing} propose using Artificial Intelligence, specifically deep learning, to optimize the performance of SAGINs by addressing key challenges such as network control, spectrum management, energy management, routing, and security. Zhang et al.\cite{zhang2022space} propose a service function chain mapping method for SAGIN that improves resource allocation and reduces latency, achieving better performance by using delay prediction and a k-shortest path algorithm. Researchers in \cite{jia2021toward} propose an optimization framework using Benders decomposition and a time-expanding graph model to enhance data collection and transmission in space-air-ground integrated networks by cooperating with high-altitude platforms and low earth orbit satellites. Existing research primarily concentrates on optimization within the confines of a limited ground station network. In contrast, our study introduces SkyGS, a novel approach that significantly enhances the existing network by integrating ground stations from various providers. This integration expands the network's reach and addresses the fundamental issue of ground station scarcity at its source. Our paper offers a unique perspective, proposing an innovative solution to a long-standing limitation in satellite communication networks.
 
\noindent\textbf{Sky computing:}
Sky Computing is merely cloud computing mediated by an intercloud broker\cite{chasins2022sky}. Research has significantly advanced this field\cite{stoica2021cloud, yang2023skypilot, jain2023skyplane, petcu2011building}. The researchers of \cite{stoica2021cloud} highlight the economic barriers in Sky Computing and propose reciprocal peering as a fundamental step. Researchers in \cite{yang2023skypilot} propose to create a fine-grained two-sided market via an intercloud broker named SkyPilot, which enables users to seamlessly run their batch jobs across clouds. In addition, researchers in \cite{jain2023skyplane} focus on the data transfer across the cloud and propose to create a system for bulk data transfer named SkyPlane through identifying overlay paths. While existing research in Sky Computing has primarily focused on integrating various cloud providers, this paper proposes a significant extension to this paradigm. We introduce a novel concept that integrates ground stations across different providers. This unique integration not only broadens the scope of Sky Computing but also addresses critical challenges in space networks.

\section{System Model and Problem Formulation}
\begin{figure}[t]
\centerline{\includegraphics[width=0.65\textwidth]{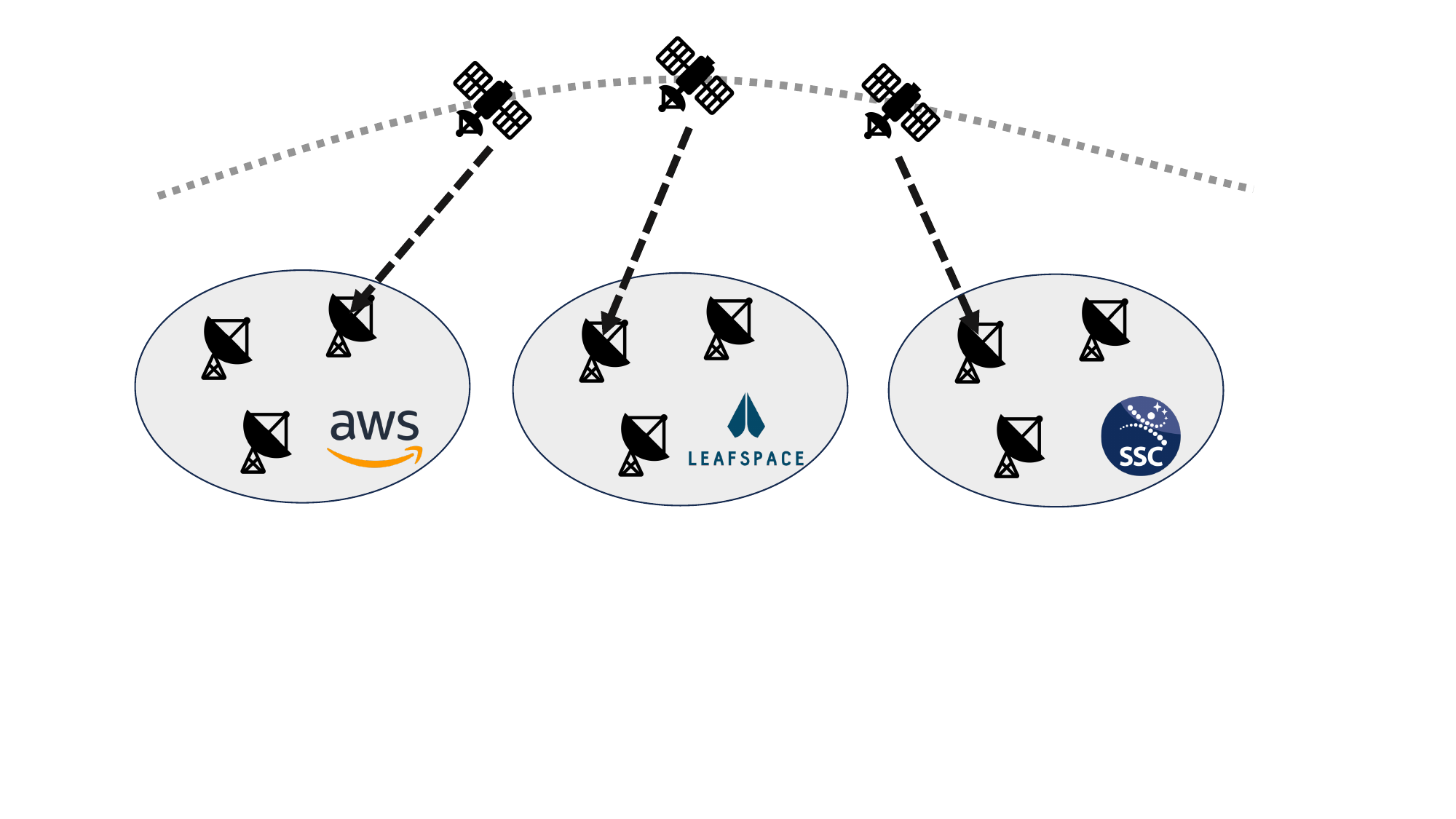}}
\caption{System overview}
\label{system overview}
\end{figure}
\subsection{System model}
Our SkyGS system is based on LEO satellite constellations, where each satellite' observed data is first transmitted to a ground station through GSL for reception, and then to a data center through a terrestrial network for processing, as depicted in Figure \ref{system overview}. The processing involves tasks like correcting atmospheric and surface distortions, orthorectifying images, and generating metadata. During the whole process, it would bring some costs, including the rental fees for the ground station and the computational cost at the data center. Furthermore, it would incur non-negligible latency including the queuing latency that data waits for transmission at the satellite, the transmission latency between the satellite, the ground station, and the data center, as well as the computation latency incurred by the data processing at the data center. In SkyGS, time is assumed to be slotted and denoted as $\mathcal{T}=\{0,1,2,..., T\}$. The length of each slot is identical and denoted as $\tau$. It is assumed that the system status keeps static during a time slot. We denote the set of satellites as $\mathcal{S}$, the set of ground stations as $\mathcal{G}$, the set of ground stations' antennas that receive data from satellites as $\mathcal{A}$, and the set of data centers as $\mathcal{D}$. We represent the number of antennas at ground station $g$ as $\psi_g$, indicating that ground station $g$ can simultaneously communicate with $\psi_g$ satellites. Moreover, one data center can simultaneously process data from any ground station without limit. We assume the locations of satellites can be predicted using orbital parameters \cite{vasisht2021l2d2} and denote the set of satellites within the view of ground stations during each time slot as $S(t)$. The set of available ground stations for satellite $s$ is denoted as $\mathcal{G}^s(t)$. The set of all antennas in $\mathcal{G}^s(t)$ is denoted as $\mathcal{A}^s(t)$. SkyGS is operated by a centralized broker in the cloud in real-time, following the sky computing paradigm. During each time slot, satellite status information, such as data backlog and downlink rate, being probed in real-time, are transmitted to the broker. Availability of ground stations and data centers is requested by the broker from corresponding providers. Satellites request data downlink for all visible ground stations at each time slot, and the ground stations selectively respond to these requests based on the scheduling results.
 
\subsection{Satellite data backlog}
Each satellite $s$ maintains a time-varying sequence of data as it collects and occasionally downlinks data to ground stations. The amount of data backlog for satellite $s$ at time slot $t$ is denoted as $D_s(t)$ and is modeled using a discrete-time queuing system. $D_s^{i}(t)$ is denoted as the amount of new data that arrives at time $t$, while $D_s^{o}(t)$ represents the amount of data that satellite $s$ can downlink at time $t$. 
 
Let $D_{gd}^s(t)$ represent the amount of data that can be downlinked at maximal from satellite $s$ to ground station $g$, and subsequently to data center $d$, during time slot $t$. $D_{gd}^s(t)$ is dependent on the link quality between satellite $s$ and ground station $g$, as well as the contact time $\tau$, calculated as: 
\begin{equation}
    D_{gd}^s(t) = x_{gd}^s(t) R_{s,g}(t) \tau
\end{equation}
where $x_{gd}^s(t)$ is a binary indicator, which equals 1 when satellite $s$ selects ground station $g$ and data center $d$ for data reception and processing at time slot $t$. $R_{s,g}(t)$ is the downlink data rate from satellite $s$ to ground station $g$ during time slot $t$, which is probed in real-time at the beginning of each time slot. Then based on $D_{gd}^s(t)$, we can express $D_s^{o}(t)$ as follows:
\begin{equation}
    D_s^{o}(t) = \sum\limits_{g \in \mathcal{G}^s(t)} \sum\limits_{d \in \mathcal{D}} D_{gd}^s(t)
\end{equation}

Since satellite $s$ can transmit at most $D_s(t)$ of data given its limited backlog amount, we then introduce the variable $\tilde{D}_s^{o}$ as the actual amount of data downlinked by satellite $s$ during time slot $t$, calculated as: 
\begin{equation}
\label{D_out_actual}
    \tilde{D}_s^{o}(t) = \min\{D_s^{o}(t),D_s(t)\}
\end{equation} 
 
Subsequently, the evolution of $D_s(t)$ over time is governed by a dynamic equation as follows:
\begin{equation}
\label{dynamic equation for satellite backlog}
    D_s(t+1) = D_s(t)- \tilde{D}_s^{o}(t) + D_s^{i}(t)
\end{equation}

\subsection{Latency modelling}
Latency in our system is defined as the time elapsed from when data is received by satellites to when they finish processing at a data center. This latency comprises five distinct components: (1) queuing latency that data waits for transmission on the satellite,  (2) propagation latency between satellites and ground stations, (3) transmission latency between satellites and ground stations, (4) transmission latency between ground stations and data centers, and (5) computation latency incurred by data processing in data centers. Given that LEO satellites operate at altitudes below 1000 km, the associated propagation latency between satellites and ground stations is a few milliseconds, which is negligible compared with the other parts of latency. Therefore, we exclude propagation latency from our analysis. The calculation of latency is based on the unit of 1MB of data. Let $N^s(t)$ denote the unit data set downlinked by satellite $s$ during time slot $t$. The cardinality of $N^s(t)$ is equal to $\tilde{D}_s^{o}$.

\textit{Queuing latency:} We define $L_{q}^s(n, t)$ as the queuing latency of downlinked data unit $n$ from satellite $s$, representing the waiting time of the data unit $n$ from being captured to the commencement of transmission to a ground station at time $t$. Let $L_{q}^s(t)$ denote the sum of queuing latency for all data units in $N_s(t)$. It can be calculated as:
\begin{equation}
    L_{q}^s(t) = \sum\limits_{n \in N^s(t)} L_{q}^s(n, t)
\end{equation}

\textit{Transmission latency between satellites and ground stations:} We define $L_{t1}^s(t)$ as the sum of transmission latency for all data units in $N_s(t)$ between satellite $s$ and a selected ground station $g$ at time $t$. This can be calculated as:
\begin{equation}
   L_{t1}^s(t) = \tilde{D}_s^{o}(t) / R_{s,g}(t)
\end{equation}

\textit{Transmission latency between ground stations and data centers:} We define $L_{t2}^s(t)$ as the sum of transmission latency for all data units in $N_s(t)$ downlinked by satellite $s$ at time $t$ between the selected ground station $g$ and data center $d$. Let $R_{g,d}$ denote the transmission data rate from ground station $g$ to data center $d$. $L_{t2}^s(t)$ can be calculated as:
\begin{equation} 
    L_{t2}^s(t) = \tilde{D}_s^{o}(t) / R_{g,d}
\end{equation}

\textit{Computation latency:} We define $L_{c}^s(t)$ as the computation latency for all data units in $N_s(t)$ downlinked by satellite $s$ at time $t$, incurred by the data processing at data center $d$. It is assumed to be proportional to the data amount $\tilde{D}_s^{o}(t)$.
 
Finally, the total latency for downlinked data $\tilde{D}_s^{o}(t)$ from satellite $s$ at time slot $t$, denoted as $L^s(t)$, can be calculated by summing all the above latency:
\begin{equation}
\label{Latency Definition}
L^s(t) = L_q^s(t) + L_{t1}^s(t) + L_{t2}^s(t) + L_{c}^s(t)
\end{equation}

\subsection{Cost modelling}
We consider the total monetary cost from two parts: the rental cost of ground stations to receive the data from satellites, and the computational cost of data centers for data processing. 
 
\textit{Ground Station Rental Cost:} 
We adopt an on-demand pay-as-we-go pricing model as in \cite{AWSpricing}, where the rental cost of ground stations is based on the request period and the number of concurrent antennas in operation. We denote the per-time-slot rental cost of ground station $g$ for turning up one antenna as $P_{g}$. Let $C_r^s(t)$ be the ground station's rental cost for satellite $s$ at time $t$, then it equals $P_{g}$ if $s$ chooses to downlink data to ground station $g$, and 0 if $s$ chooses to backlog the data at $t$. 

\textit{Computational Cost:} 
We consider the computational cost at data center $d$ per time slot as a constant, denoted as $P_d$. Assume satellite $s$ chooses data center $d$ to process the data downlinked at time $t$, then the computational cost for this amount of data, denoted as $C_c^s(t)$ is calculated as:
\begin{equation}
C_c^s(t) = P_dL_c^s(t) 
\end{equation}

Hence, the total monetary cost of satellite $s$ for data downlinked at time $t$ can be calculated as
\begin{equation}
\label{C_{gd}^s(t)}
C^s(t) = C_r^s(t) + C_c^s(t)
\end{equation}

Then we get the total monetary cost at time $t$ as: 
\begin{equation}
    C(t)  = \sum\limits_{s \in \mathcal{S}(t)} C^s(t)
\end{equation}

\subsection{Problem Formulation}
Given satellites' limited onboard storage capacity, it is critical to mitigate the risk of data overflow and ensure controllable data latency. We thereby enforce the time-varying data backlog queue of satellite $s$ (i.e. $D_s(t)$) to be mean rate stable, meaning that the average arrival rate of data to the queue does not exceed the average departure rate of data from the queue. This prevents the queue length from growing indefinitely, ensuring the system can operate stably over the long term. Mathematically, it could be expressed as:
\begin{equation}
\label{Queue mean rate stable}
    \lim_{T \to \infty} \mathbb{E}[D_s(T)]/T = 0\quad \forall s \in \mathcal{S} 
\end{equation}

Furthermore, given the urgent need for rapid data transmission and processing, especially in scenarios such as disaster alarms, it is imperative to establish an upper limit for data latency. Since satellites will operate in orbit over the years, we establish a long-term latency threshold for the overall downlinked data. This latency constraint could be expressed as:
\begin{equation} 
\label{latency_simplied}
    \lim\limits_{T \to \infty} \frac{1}{T} \sum\limits_{t=0}^{T-1} \sum\limits_{s \in \mathcal{S}(t)} \mathbb{E}[ L^s(t) -\xi \;\tilde{D}_s^{o}(t) ] \leq 0
\end{equation}
where $\xi$ is the per-data-unit latency threshold such that the overall sum of per-data-unit latency does not exceed $\xi$ times the overall amount of downlinked data. 

We aim to optimize long-term performance within a predefined long-term latency threshold. Our objective for each time slot is to identify the appropriate selection policy for satellite, ground station, and data center pairs that minimize the system cost, adhering to the above constraints. Accordingly, we formulate the problem as follows:
\begin{align} \label{P1}
\textbf{P1.} \quad & \min_{\mathbf{x}} \lim_{T \to \infty} \frac{1}{T} \sum\limits_{t=0}^{T-1} \mathbb{E}[C(t)]\\
 \nonumber \text{s.t.} \quad & \eqref{Queue mean rate stable},\eqref{latency_simplied} \\
 \label{c_15}
 &\sum\limits_{g \in \mathcal{G}^s(t)}\sum\limits_{d \in \mathcal{D}} x_{gd}^s(t) \leq 1,\quad \forall s\in\mathcal{S}, \forall t\in\mathcal{T}  \\
  \label{c_16}
 &\sum\limits_{g \not\in \mathcal{G}^s(t)}\sum\limits_{d \in \mathcal{D}} x_{gd}^s(t) = 0,\quad \forall s\in\mathcal{S},\forall t\in\mathcal{T} \\
  \label{c_17}
 &\sum\limits_{s \in \mathcal{S}}\sum\limits_{d \in \mathcal{D}} x_{gd}^s(t) \leq \psi_g, \quad \forall g\in\mathcal{G}, \forall t\in\mathcal{T} \\
  \label{c_18}
 &x_{gd}^s(t) \in  \{0,1\}, \quad  \forall s\in\mathcal{S},\forall g\in\mathcal{G},\forall d\in\mathcal{D},\forall t\in\mathcal{T}
\end{align}

where $\mathbf{x} = [\mathbf{x}(t)]_t$, $\mathbf{x}(t) = [x_{gd}^s(t)]_{s \in \mathcal{S}, g \in \mathcal{G}, d \in \mathcal{D}}$. Constraint \eqref{c_15} ensures that each satellite at each time slot could only select a ground station and a data center for reception and computation, respectively. Constraint \eqref{c_16} ensures a satellite can not downlink data to a ground station that is out of its current visible range.  Constraint \eqref{c_17} ensures that the number of satellites with which a ground station can concurrently communicate cannot exceed the number of antennas it possesses.

\section{Algorithm Design and Theoretical Analysis}
\subsection{Algorithm Design}
 Due to the dynamic properties of the system (e.g. fluctuation of GSL quality), it is hard to find the optimal solution without global information. Lyapunov optimization is a method that only requires knowledge of the current system state without requiring any prior knowledge. It leverages control theory to ensure the stability of dynamic systems while simultaneously optimizing specific performance objectives \cite{definition-of-Lyapunov-Optimization}. Typically, this dynamic system is represented by a sequence of actual and virtual queues, which needs to be stabilized by minimizing the drift of a quadratic Lyapunov function.

To apply the Lyapunov optimization framework, we first transform \textbf{P1} into a problem of minimizing the Lyapunov drift-plus-penalty function. We begin with transforming Constraint \eqref{latency_simplied} into a queue stability problem by constructing a virtual queue. We define $\phi^s(t)$ as the latency exceeding the long-term data latency threshold $\xi$ for data originating from satellite $s$.
\begin{equation}
    \phi^s(t) =  L^s(t) - \xi \tilde{D}_s^{o}(t)
    \label{phi}
\end{equation}

Then the total latency beyond threshold is $\phi(t) = \sum_{s \in \mathcal{S}}\phi^s(t)$. $Q(t)$ is defined as the virtual queue backlog of data latency beyond the long-term latency threshold $\xi$ and assumes that the initial queue backlog is 0 (i.e., $Q(0)=0$). 
\begin{equation}
Q(t+1) = \max\Big\{Q(t) + \phi(t), 0\Big\}
\label{Q(t)}
\end{equation}

Intuitively, the value of $Q(t)$ can be regarded as an evaluation criterion to assess the data latency condition. A high value of $Q(t)$ suggests that the latency has far exceeded the predefined threshold. The stability of $Q(t)$ is equivalent to adhering to the time-average latency constraint \eqref{latency_simplied}. If the virtual queue $Q(t)$ is mean rate stable, the time average latency constraint \eqref{latency_simplied} can also be satisfied. That is,
\begin{equation} \label{constraint_to_queue_stability}
    \lim_{T \to \infty} \mathbb{E}[Q(T)]/T = 0 \Rightarrow \lim_{T \to \infty}\frac{1}{T}\sum\limits_{t=0}^{T-1}\mathbb{E}[\phi(t)] \leq 0
\end{equation}

Please see the Appendix \ref{proof_of_constraint_to_queue_stability} for the proof detail.

 Since virtual queue $Q(t)$ and satellite backlog queues $D_s(t), s \in \mathcal{S}$ should be mean rate stable, we concatenate them as $\bold{\bold{\Theta}}(t) = [Q(t), D_1(t), D_2(t), .., D_{|\mathcal{S}|}(t)]$. We then define the Lyapunov function as follows: 
\begin{equation} 
    L(\bold{\bold{\Theta}}(t)) \triangleq \frac{1}{2}Q(t)^2 + \frac{1}{2}\sum\limits_{s=1}^{|\mathcal{S}|} D_s(t)^2
\end{equation}
which is a non-negative scalar measure of queue backlog. Then we introduce \textit{one-slot conditional Lyapunov drift} $\Delta(\bold{\bold{\Theta}}(t))$, which is the expected change in the Lyapunov function over one slot, given that the current state at slot $t$ is $\bold{\Theta}(t)$:
\begin{equation}
    \Delta(\bold{\bold{\Theta}}(t)) \triangleq \mathbb{E}\left [L(\bold{\Theta}(t+1)) - L(\bold{\Theta}(t)) | \bold{\Theta}(t)\right ]
\end{equation}

We could transform \textbf{P1} to \textbf{P2}, which minimizes the Lyapunov drift-plus-penalty function for the joint concern of queue stability and system cost optimization.
\begin{align}
\textbf{P2:} &\label{p2} \quad \min_{\mathbf{x}(t)} \Delta(\bold{\Theta}(t)) + V\mathbb{E}[C(t)|\bold{\Theta}(t)]\\
s.t. & \quad \eqref{c_15}\eqref{c_16}\eqref{c_17}\eqref{c_18} \nonumber
\end{align}
where $V$ is a non-negative tunable parameter reflecting the trade-off between backlog reduction and penalty minimization.

Rather than seeking a direct solution, which requires future information, the problem \textbf{P2} can be solved by minimizing its upper bound. The following Theorem gives the upper bound of the drift-plus-penalty function.
\begin{theorem} \label{theorem 1}
Under any control algorithm, the drift-plus-penalty expression has the following upper bound for all $t$, all possible values of $\bold{\Theta}(t)$, and all parameter $V \geq 0$:
\begin{align} \label{theorem 1 equ}
    &\Delta(\bold{\Theta}(t)) + V\mathbb{E} \left [ C(t)|\bold{\Theta}(t) \right ] \leq B + V\mathbb{E}\left [C(t)|\bold{\Theta}(t)\right ] + \\
    & \sum\limits_{s=1}^{|\mathcal{S}|}D_s(t)\mathbb{E}\left [D_s^{i}(t)-D_s^{o}(t)|\bold{\Theta}(t)\right ] + Q(t)\mathbb{E}\left [ \phi(t)|\bold{\Theta}(t)\right ] \nonumber
\end{align}
 where $B$ is defined as a positive constant that satisfies the following for all $t$:
\begin{equation}\label{B} 
    B \geq \frac{1}{2}\sum\limits_{s=1}^{|\textit{S}|}\left [D_{s,max}^{i}(t)^2+D_{s,max}^{o}(t)^2\right ] + \frac{1}{2}\phi_{max}(t)^2 
\end{equation}
Such a constant $B$ exists because $D_s^{i}(t)$,  $D_s^{o}(t)$ and $\phi(t)$ are assumed to be bounded. $D_{s,max}^{i}(t) = \max_{\mathbf{x}(t)}D_s^{i}(t)$, $D_{s,max}^{o}(t)=\max_{\mathbf{x}(t)}D_s^{o}(t)$, and $\phi_{max}(t)=\max_{\mathbf{x}(t)}\phi_(t)$.
\end{theorem} 
Please see the Appendix \ref{Appendix T1} for the proof detail.

At every time slot $t$, by observing the current queue status $\bold{\Theta}(t)$ and system information(e.g., GSL quality), we minimize the upper bound of the drift-plus-penalty function. Thus, we can transform problem \textbf{P2} to \textbf{P3} as follows:
\begin{align}
\textbf{P3:} &\label{p3} \; \min_{\mathbf{x}(t)}  VC(t)+ \sum\limits_{s=1}^{|\mathcal{S}|}D_s(t)\left(D_s^{i}(t)-D_s^{o}(t)\right) + Q(t)\phi(t) \\
s.t. & \quad \eqref{c_15}\eqref{c_16}\eqref{c_17}\eqref{c_18} \nonumber
\end{align}

 To solve \textbf{P3}, the commonly used algorithm which minimize the upper bound of drift-plus-penalty function at each time slot is to enumerate all possible $\mathbf{x}(t)$
 at time $t$ and find the one leading to the optimal solution. However, the time complexity of enumeration is $\mathcal{O}(2^{|\mathcal{S}|\times |\textit{G}| \times |\textit{D}|})$, which is of exponential complexity. Therefore, developing a more efficient approach than simple enumeration is necessary. Our solution involves constructing a weighted bipartite graph and computing the minimal matching using the Hungarian algorithm.
\begin{algorithm} 
	\caption{Drift-plus-penalty algorithm} 
	\label{drift_plus_penalty} 
	\begin{algorithmic}[1]
        \STATE Initialization: \text{Set initial value of all actual and virtual}
        \STATE queues zero
        \FOR{each time slot $t = 0,1,2,...,T$} {
            \STATE \textbf{Find} $\mathbf{x}(t) = \mathop{\arg\min} \textbf{P3}$ \\
            \STATE \textbf{Update} \text{all actual and virtual queues} 
        }
        \ENDFOR
	\end{algorithmic} 
\end{algorithm}

To be specific, for each time slot $t$, we construct a bipartite graph $G=(\mathcal{S}, \mathcal{A}\cup \mathcal{S}, E)$ in the broker, where the left subset represents all the satellites $\mathcal{S}$ and the right subset denotes all the antennas including $|\mathcal{S}|$ virtual antennas alongside actual antennas $\mathcal{A}$. Each virtual antenna is linked to a corresponding satellite with an edge, symbolizing scenarios where the satellite fails to downlink data due to reasons such as not being within the visual range of the ground stations or contention between satellites. Furthermore, an edge $(s, a)$ is created for $s \in \mathcal{S}(t), a \in \mathcal{A}^s(t)$. For each edge, the satellite and ground station are determined. To determine the weight of each edge, we need to select the corresponding data center. Eq. \eqref{p3} can be decomposed based on satellites. For minimization, for each edge $(s,a)$ we we select data center $d$ satisfying 
\begin{equation}
\label{find dc for each edge}
    \min_{d} VC^s(t) + D_s(t)\left(D_s^{i}(t)-D_s^{o}(t)\right) + Q(t)\phi^s(t)
\end{equation}
We then apply the Hungarian algorithm to find the minimal matching. By summing up all the resulting edges, we obtain the minimization value of Eq. \eqref{p3} at each time slot.  

Determining the weight of all edges takes $O(|E||\mathcal{D}|)$ since each edge needs to find the minimum weight value within $|\mathcal{D}|$ data centers, as specified in Eq.\eqref{find dc for each edge}, while Hungarian algorithm takes $O((|\mathcal{S}|+|\mathcal{A}|)^3)$. Hence, the complexity of determining the optimal $\mathbf{x}(t)$ is $O(|E||\mathcal{D}| + (|\mathcal{S}|+|\mathcal{A}|)^3)$.

\subsection{Theoretical Analysis}
We then analyze the performance of drift-plus-penalty algorithm theoretically and show its gap from the optimal solution. 
\begin{theorem} \label{theorem 2}
Suppose \textbf{P1} has a feasible solution, the time-averaged expected costs achieved by our algorithm have a bounded gap with the optimal optimization costs, which can be described below:
\begin{equation}
\label{cost_upper_bound}
    \lim_{T \to \infty} \frac{1}{T} \sum_{t=0}^{T-1} \mathbb{E}\left[C(t)\right] \leq C^* + \frac{B}{V} 
\end{equation}
where $C^*$ is the infimum time-averaged cost under any policy meeting the constraints, and $B$ is defined in Constraint \eqref{B}. 
\end{theorem}
\begin{theorem} \label{theorem 3}
The sum of the backlog of the actual and virtual queues is stable, which is presented as follows:
\begin{equation*}
\label{queue backlog upper bound}
    \lim_{T\to\infty}\frac{1}{T}\left\{\sum_{t=0}^{T-1}\sum_{s=1}^{\mathcal{|S|}} \mathbb{E}[D_s(t)]+\sum_{t=0}^{T-1}\mathbb{E}[Q(t)] \right\} \leq \frac{B'}{\epsilon}
\end{equation*}
where $\epsilon > 0, B' = B + V(C_{max}-C_{min})$, $C_{max}$ and $C_{min}$ are the maximum and minimum system costs in all selection $\mathbf{x}(t)$ respectively.
\end{theorem}
Please see the \ref{Appendix T2} and \ref{Appendix T3} for the proof detail.

The above theorems reveal that the time-average costs achieved by our algorithm are within a constant gap $B / V$ from the optimal costs. Furthermore, the time-average queue backlog is also bounded by a fixed value, which increases linearly with the parameter $V$.

\section{Experiment}
In this section, we conduct a series of experiments to evaluate the time-averaged cost performance under the long-term latency threshold and compare it with several alternative strategies. 
 
\subsection{Simulation Setup}
\begin{figure}[t!]
    \centering
    \subfigure[Ground station]{
    \includegraphics[width = 0.22\textwidth]{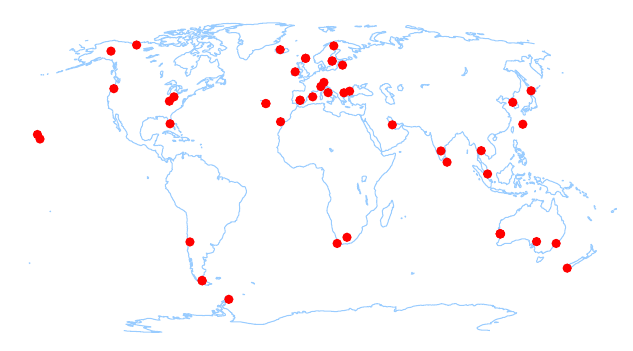}
        \label{fig:gs_location}
    }
    \subfigure[Data center]{
\includegraphics[width = 0.22\textwidth]{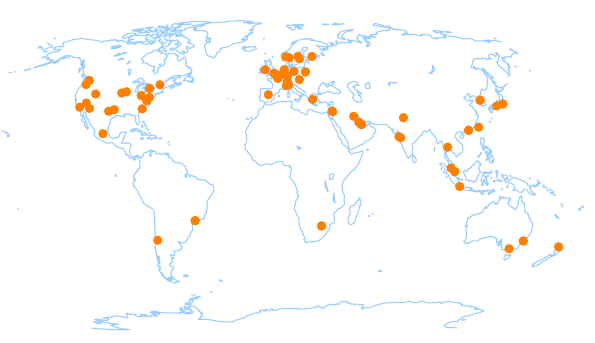}
        \label{fig:dc_location}
    }
    \caption{Ground station and data center locations}
    \label{fig:gs_dc}
\end{figure}

\textbf{Constellation-ground station-data center simulation:} Our satellite dataset originates from the Dove constellation, operated by Planet Inc\cite{planet}. This constellation consists of Earth observation satellites positioned at an altitude of 475 km in Sun Synchronous Orbits (SSO)\cite{PlanetNASABrownBag}. 
We conducted simulations of 153 satellites using Pyorbital\cite{Pyorbital}, leveraging Two Line Element set (TLE) data sourced from Celestrak\cite{Celestrak} to estimate satellite positions relative to Earth. To federate GSaaS, 48 ground stations have been employed from prominent providers such as AWS\cite{AWS}, Leaf Space\cite{leafspaceservices}, and SSC\cite{ssc_ground_station}, as depicted in Fig. \ref{fig:gs_location}. The emulation of 109 data centers is modeled on the geographical distribution of cloud sites from major providers Amazon AWS\cite{aws_dc}, Microsoft Azure\cite{azure_dc}, and Google Cloud Platform\cite{gcp_dc}, which accounts for $67\%$ of the global cloud infrastructure market\cite{cloudzero}.

\textbf{Data Collection Mode:} 
According to work \cite{tao2023transmitting, foster2015orbit}, the Dove satellite constellation continuously images while over land with reasonable solar illumination with each satellite collecting 1 TB data daily. Hence, we assume each satellite collects $[0.9, 1.1]$ TB data per day.
 
\textbf{Other hyperparameter settings:} 
 We employed a discrete-event simulation with a time granularity of one-minute intervals for one day. According to Devaraj et al. \cite{devaraj2019planet}, the peak downlink data rate for the Dove constellation can reach 1.6 Gbps. In each time slot, we use Pyorbital \cite{Pyorbital} to determine viable satellite-ground station pairs and estimate the available data rate between them. This data rate is scaled to a maximum of 1.6 Gbps, with random noise added to simulate weather effects. Additionally, following the methodology of Bill Tao \cite{tao2023transmitting}, we set the data rate for all connections between ground stations and data centers to 1 Gbps. Each ground station is randomly assigned 1, 2, or 3 antennas.
For antenna rental, the cost per minute is uniform within each provider. We simulate each provider's ground station costs at 18, 22, and 26 dollars per minute. The processing time for one gigabyte of image data is influenced by data center load and data type, assumed to range from 0.1 to 0.2 hours based on the actual case\cite{yan2018cloud}. Referencing the pricing models from Amazon's website, we determine data center costs. These costs vary by region, with data center usage priced between 0.5 and 1 dollar per hour.

\subsection{Benchmarks}
To demonstrate the efficiency of our algorithm, we conducted a comparative analysis against five alternative algorithms. The first strategy involves using a single provider for the selection of ground stations and data centers. The remaining four strategies employ multiple providers for these selections.
\begin{itemize}
\item \textbf{Single-Provider-Greedy (SG):} This approach restricts selection to a single provider for both ground station and data center and greedily selects cost-effective pairs of satellite, ground station, and data center for data downlink.

\item \textbf{Broker-Greedy (BG):} This algorithm greedily selects cost-effective data downlink pairs from all ground stations and data centers across diverse providers.

\item \textbf{Broker-Random (BR):} Selection of data downlink pairs occurs randomly from all ground stations and data centers.

\item \textbf{Broker-Withhold-Greedy (BWG):} Building on the BG method, this approach also greedily selects cost-effective data downlink pairs but adds the strategy of satellites withholding data until the GSL bandwidth can be fully leveraged.

\item \textbf{ILP:} Inspired by SkyPilot\cite{yang2023skypilot}, we propose using an Integer Linear Programming (ILP) solver to address our problem. This algorithm employs a broker for selection, linearizes the problem, and uses the General Linear Programming Kit (GLPK) to solve the ILP at each time slot. Since linearizing the queue stability constraint \eqref{Queue mean rate stable} and the time-average constraint \eqref{latency_simplied} is challenging,  we introduce a high-priority downlink queue. When data queuing latency from a satellite approaches the threshold $\xi$, the satellite is moved to this high-priority queue, enforcing immediate downlink.
\end{itemize}

\begin{figure*}[t!]
    \centering
    \subfigure[System cost]{
    \includegraphics[width = 0.3\textwidth]{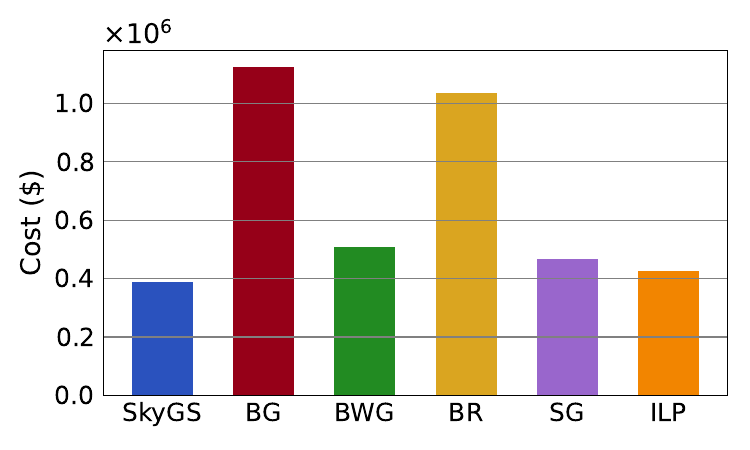}
        \label{fig:cost_bar}
    } 
    \subfigure[Latency CDF (vertical red dashed line represents latency threshold)]{
\includegraphics[width = 0.3\textwidth]{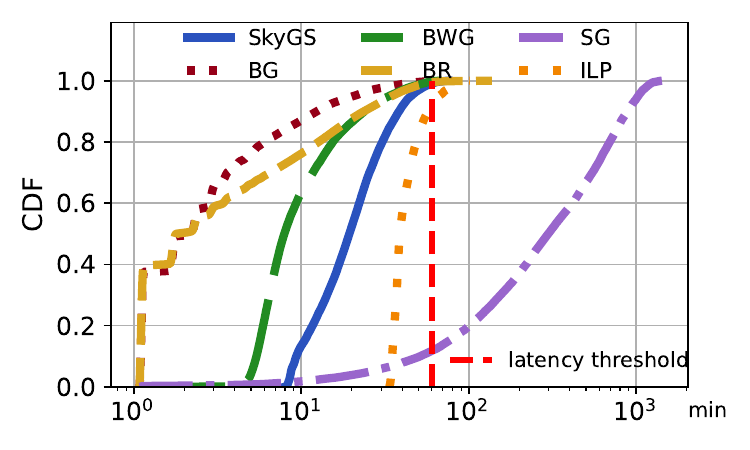}
        \label{fig:latency_cdf}
    }
    \subfigure[Average latency and latency violation ratio]{
\includegraphics[width = 0.3\textwidth]{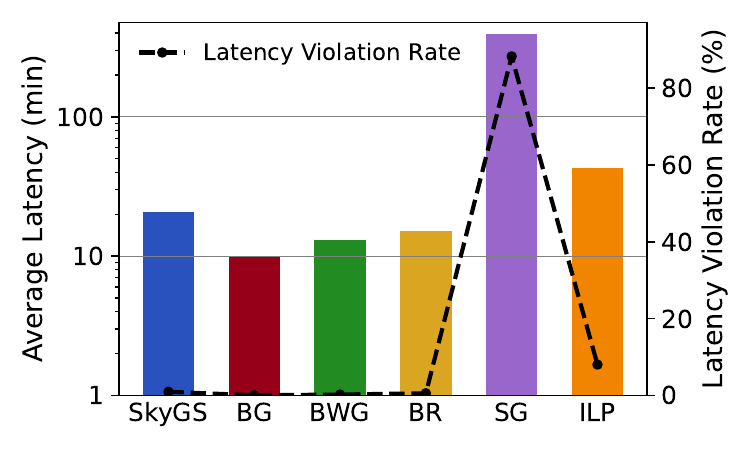}
        \label{fig:latency_bar}    }
    \caption{Comparison between different strategies}
    \label{fig:comparison}
\end{figure*}

\begin{figure*}[t!]
    \centering
    \subfigure[Impact of parameter $V$]{
    \includegraphics[width = 0.3\textwidth]{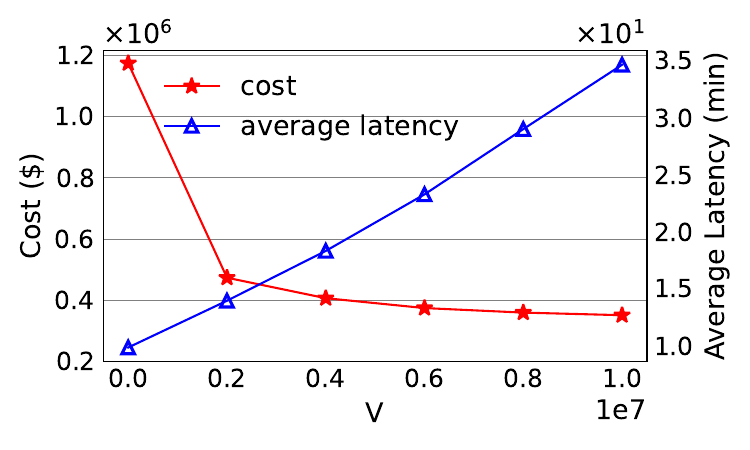}
        \label{fig:V}
    }  
    \subfigure[Impact of GSL data rate]{
\includegraphics[width = 0.3\textwidth]{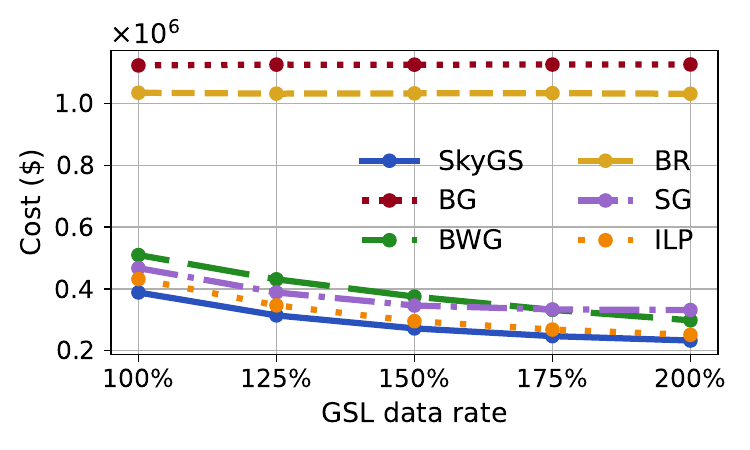}
        \label{fig:downlink_rate_cost}
    } 
    \subfigure[Impact of constellation size]{
\includegraphics[width = 0.3\textwidth]{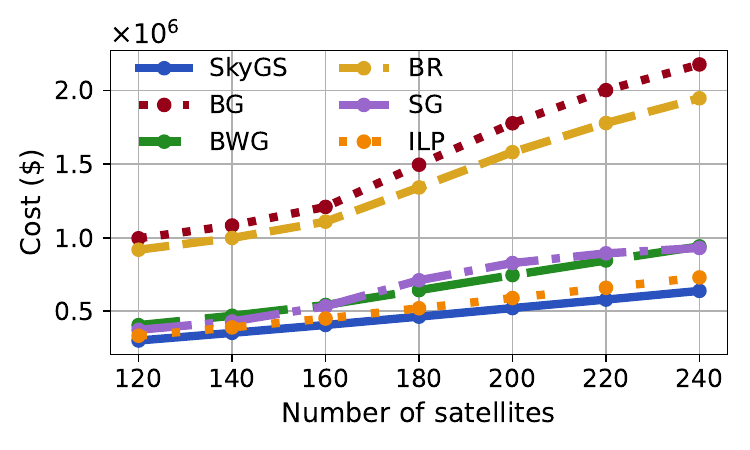}
        \label{fig:constellation_size_cost}
    } 
    \caption{Sensitivity analysis}
\end{figure*}

\subsection{Metrics}
To analyze the performance of different algorithms, we introduce the following metrics:
\begin{itemize}
\item \textbf{System cost:} Total expenditure incurred throughout the simulation.
\item \textbf{Average latency:} The average time from data capture to the end of processing for all data units.
\item \textbf{Latency threshold violation rate:} The ratio of data downlinks whose average latency exceeds the latency threshold $\xi$.
\end{itemize}

\subsection{Overall performance}
We set the average latency threshold $\xi$ to be 60 minutes to ensure a relatively low data latency. The parameter $V$ is set to 5,000,000, which minimizes system cost while maintaining the latency of most data within the latency threshold $\xi$. Since AWS provides follow-up data processing services for GSaaS, SG adopts AWS as the sole provider.
 
\textbf{System cost comparison:} We assess the costs of six strategies over 24 hours, illustrated in Fig. \ref{fig:cost_bar}. The SkyGS strategy incurs the lowest system cost compared to the alternatives. Specifically, SkyGS results in 16.9\%, 65.5\%, 62.5\%, 23.8\%, and 9.0\% cost savings, compared with SG, BG, BR, BWG, and ILP respectively. Each strategy selects the cheapest data centers for processing, given the static link between ground stations and data centers. The performance of cost savings is determined by selecting cost-effective ground stations and minimizing times of data downlink. SkyGS excels due to its unique downlink scheduling approach, which considers latency queue backlog, satellite data backlog, and cost for each time slot. This allows SkyGS to withhold data when the backlog is small and latency is acceptable, reducing the number of downlinks and lowering costs. Conversely, BG performs the worst because it greedily downlinks data without withholding, leading to underutilized GSL bandwidth and excessive downlinks. BWG, despite fully utilizing bandwidth, fails to select high-quality, cost-effective ground stations. SG performs well because it passively reduces downlinks due to limited ground station availability. ILP performs worse than SkyGS because it operates on a slot-by-slot basis without considering past and future information, while SkyGS uses queue backlogs to make more strategic and holistic scheduling decisions.

\textbf{Data latency comparison:} We then examine data latency performance. As shown in Fig. \ref{fig:latency_cdf} and \ref{fig:latency_bar}, SkyGS outperforms SG and ILP with an average latency of 20.7 minutes and a 0.95\% latency violation rate. SG shows the worst performance with an average data latency of 395.6 minutes and an 88.2\% latency threshold violation rate due to limited ground station availability, highlighting the advantage of a federated ground station network. On the other hand, BG and BWG achieve lower latency since they greedily downlink data for great concern of data latency, while SkyGS and ILP tend to withhold data, thereby increasing data latency. However, our SkyGS could achieve a small latency violation ratio ($<1\%$), validating the satisfaction of our latency constraints.

\subsection{Sensitivity analysis}

\textbf{Impact of parameter $V$:} Parameter $V$ is a control parameter that is critical in determining the trade-off between minimizing data latency and costs. Here we evaluate the impact of the parameter $V$ on both the system cost and average data latency. As demonstrated in Fig. \ref{fig:V}, an increase in $V$ leads to a marked decrease in costs, which then plateau at a consistent value, in agreement with Equ. \eqref{cost_upper_bound} from Theorem \ref{theorem 2}. Additionally, according to Theorem \ref{theorem 3}, the sum of the time-average queue backlog bound increases linearly with $V$. So, average latency also increases almost linearly with $V$.

\textbf{Impact of GSL data rate:} As spacecraft hardware variants evolve, the GSL data rate continues to increase\cite{devaraj2019planet}. We examine the impact of this increase on system costs. During the experiment, we adjust parameter $V$ and proactively move satellites into the high-priority queue ahead of time to control the latency threshold violation rate of SkyGS and ILP less than $1\%$. Fig. \ref{fig:downlink_rate_cost} shows the system costs for six algorithms under various GSL data rates. The system cost in SkyGS decreases as the downlink data rate increases, and is lower than that of all other strategies. Specially, SkyGS achieves cost savings by $45.6\%$, $52.8\%$, $57.3\%$, $60.2\%$, and $61.8\%$ compared with the average cost of other five algorithms when the GSL data rate is 1, 1.25, 1.5, 1.75, 2 times the original rate, respectively.

\textbf{Impact of constellation size}
We then compare the costs among various methodologies under the scenario of various constellation sizes. During this experiment, we also adjust parameters to control the latency violation rate of SkyGS and ILP within $1\%$. As depicted in Fig. \ref{fig:constellation_size_cost}, SkyGS consistently outperforms the other alternatives. In particular, SkyGS achieves cost savings by at least $47\%$ compared with the average cost of the other five algorithms, meaning that our proposed algorithm is scalable to the constellation size.

\section{Conclusion}
In this paper, we explore the Earth observation LEO satellite data downlink cost optimization problem with a long-term data latency threshold. During optimization, we consider both communication and computation costs, encompassing not only the transmission of satellite data but also its processing budgets. We introduce SkyGS which federates GSaaS and data centers from multiple providers, enhancing satellite data downlink opportunities and increasing the flexibility of data center selection. Utilizing the Lyapunov optimization framework, SkyGS dynamically selects pairs of satellites, ground stations, and data centers, which works effectively without requiring future information. We provide theoretical proofs demonstrating that our algorithm approaches an optimal solution within a constant gap. We conducted illustrative experiments to evaluate the effectiveness of our SkyGS algorithm. To be specific, compared to alternative approaches, SkyGS saves system costs by up to $63\%$ and reduces average data latency by up to 95\%.

\appendix
\subsection{Proof of Equ. \eqref{constraint_to_queue_stability}} \label{proof_of_constraint_to_queue_stability}
According to \eqref{Q(t)}, we have $Q(t+1) \geq Q(t) + \phi(t)$, sum over the time slots $t \in \{0,1,...,T-1\}$ and divide both sides by $T$, we can get
\begin{equation}
    (1/T)\sum\nolimits_{t=0}^{T-1} \phi(t) \leq (Q(T) - Q(0))/T
\end{equation}
As $Q(0) = 0$, taking expectations of the above and taking $T \to \infty$ shows
\begin{equation}
    \lim_{T \to \infty}(1/T)\sum\nolimits_{t=0}^{T-1}\mathbb{E}[\phi(t)] \leq \lim_{T \to \infty}\mathbb{E}[Q(T)] / T
\end{equation}
Thus, if $Q(t)$ is mean rate stable, the right-hand-side of the above inequality is 0 and so:
\begin{equation}
    \lim_{T \to \infty}(1/T)\sum\nolimits_{t=0}^{T-1}\mathbb{E}[\phi(t)] \leq 0
\end{equation}   
 
\subsection{Proof of Theorem 1} \label{Appendix T1} 
Squaring the actual queue update Equ. \eqref{dynamic equation for satellite backlog} and using the fact that $\max [a-b, 0]^2 \leq (a-b)^2$ yields:
\begin{align*}
    & [D_s(t+1)^2 - D_s(t)^2]/2 \leq D_s(t)[D_s^{i}(t) - D_s^{o}(t)] +\\
    & [D_s^{i}(t)^2 + D_s^{o}(t)^2]/2 - \tilde{D}_s^{o}(t)D_s^{i}(t) 
\end{align*}
Similarly, $[Q(t+1)^2-Q(t)^2]/2 \leq \phi(t)^2/2 + Q(t)\phi(t)$. Sum up and take conditional expectations,
\begin{align} 
    &\Delta(\bold{\Theta}(t)) + V\mathbb{E} \left [ C(t)|\bold{\Theta}(t) \right ] \leq B + V\mathbb{E}\left [C(t)|\bold{\Theta}(t)\right ] + \nonumber\\
    & \sum\nolimits_{s=1}^{|\mathcal{S}|}D_s(t)\mathbb{E}\left [D_s^{i}(t)-D_s^{o}(t)|\bold{\Theta}(t)\right ] + Q(t)\mathbb{E}\left [ \phi(t)|\bold{\Theta}(t)\right ] \nonumber
\end{align}
where B is set as a positive constant that satisfies the following for all $t$:
\begin{equation*}
    B \geq \frac{1}{2}\sum\nolimits_{\smash{s=1}}^{\smash{|\textit{S}|}}[D_{s,max}^{i}(t)^2+D_{s,max}^{o}(t)^2] + \frac{1}{2}\phi_{max}(t)^2\
\end{equation*}

\subsection{Proof of Theorem 2} \label{Appendix T2}
First, we define the $\omega$-only policy, which observes the random events such as GSL quality for each slot $t$ and independently choose $\mathbf{x}(t)$ as a pure (possibly randomized) function of the observed random events. $\omega$-only policy is independent of $\bold{\Theta}(t)$.
\begin{lemma} \label{lemma 1}
    Suppose the problem \textbf{P1} is feasible, then there exists an optimal $\omega$-only policy. During each time slot, we select $\mathbf{x^*}(t)$ under optimal $\omega$-only policy, then
    \begin{align}
        &\mathbb{E}[C^*(t)] = C^*, \quad \mathbb{E}[\phi^*(t)] \leq 0\\
        &\mathbb{E}[D_s^{i*}(t) - D_s^{o*}(t)]\leq 0 \quad \forall s \in \mathcal{S}
    \end{align}
    where $C^*(t)$, $\phi^*(t)$, $D_s^{i*}(t)$, $D_s^{o*}(t)$ are values under $\mathbf{x^*}(t)$ and $C^*$ is the infimum time average cost under any policy meeting the constraints.
\end{lemma} 
\noindent According to Theorem \ref{theorem 1}, we have
\begin{align}
    &L(\bold{\Theta}(t+1)) - L(\bold{\Theta}(t)) +VC(t) \nonumber\\
    \leq & B + VC(t) + \sum\nolimits_{\smash{s=1}}^{\smash{|\textit{S}|}} D_s(t)[D_s^{i}(t) - D_s^{o}(t)] + Q(t)\phi(t) \nonumber\\
\leq &B + VC^*(t) + \sum\nolimits_{\smash{s=1}}^{\smash{|\textit{S}|}} D_s(t)[D_s^{i*}(t) - D_s^{o*}(t)] + Q(t)\phi^*(t)\nonumber
\end{align}
Here $C(t)$, $\phi(t)$, $D_s^{i}(t)$, $D_s^{o}(t)$ are values under 
$\mathbf{x}(t)$ resulted from Equ. \eqref{p3}. Take the conditional expectation and apply Lemma \ref{lemma 1}, we have
\begin{align} 
    &\Delta(\bold{\Theta}(t)) + V\mathbb{E} \left [ C(t)|\bold{\Theta}(t) \right ] \leq B + V\mathbb{E}\left [C^*(t)\right ]\\
    & + \sum\nolimits_{s=1}^{|\mathcal{S}|}D_s(t)\mathbb{E}\left[D_s^{i*}(t)-D_s^{o*}(t)\right ] + Q(t)\mathbb{E}\left [ \phi^*(t)\right ] \nonumber \\
    &  \leq B + VC^*
\end{align}

Take the expectation and sum up over the time slot $t \in \{0,1,2,...T-1\}$, we can obtain
\begin{equation*}
    \mathbb{E}\left[L(\bold{\Theta}(T)) \right] - \mathbb{E}\left[L(\bold{\Theta}(0)) \right] + V\sum\nolimits_{t=0}^{T-1}\mathbb{E}\left [ C(t)\right] \leq BT + TVC^*
\end{equation*}
Since $L(\bold{\Theta}(0)) = 0, L(\bold{\Theta}(T)) \geq 0$, divide both sides by $VT$ and take $T \to \infty$:
\begin{equation*}
    \lim_{T \to \infty} (1/T) \sum\nolimits_{t=0}^{T-1} \mathbb{E}\left[C(t)\right] \leq C^* + \frac{B}{V}
\end{equation*}
 
\subsection{Proof of Theorem 3} \label{Appendix T3} 
We assume that there exists one $\omega$-only policy $\mathbf{x^{\bullet}}$ that satisfies:
\begin{equation*}
    {\exists} \epsilon>0, \mathbb{E}[\phi^{\bullet}(t)] \leq - \epsilon, \mathbb{E}\left[ D_s^{i\bullet}(t) - D_s^{o\bullet}(t)\right] \leq - \epsilon, \forall s\in \mathcal{S}
\end{equation*}

In addition, we assume that $C(t)$ is bounded, i.e.,
\begin{equation*}
    {\exists} C_{min}, C_{max} \in \mathbb{R}, \forall t,\mathbf{x}(t), C_{min} \leq C(t) \leq C_{max} 
\end{equation*}
\noindent According to Theorem \ref{theorem 1} and Equ. \eqref{p3}, we have
\begin{align}
    &L(\bold{\Theta}(t+1)) - L(\bold{\Theta}(t)) +VC(t) \leq B + VC^\bullet(t)\nonumber\\ 
    & + \sum\nolimits_{\smash{s=1}}^{\smash{|\textit{S}|}} D_s(t)[D_s^{i\bullet}(t) - D_s^{o\bullet}(t)] + Q(t)\phi^\bullet(t)\nonumber
\end{align}
Take the conditional expectation, we can obtain
\begin{align*}
    &\Delta(\bold{\Theta}(t)) + VC_{min} \leq \Delta(\bold{\Theta}(t)) + V\mathbb{E} \left [ C(t)|\bold{\Theta}(t) \right ]\\
    &\leq B + VC_{max} +\sum\nolimits_{\smash{s=1}}^{\smash{|\textit{S}|}} D_s(t)\mathbb{E}[D_s^{i\bullet}(t) - D_s^{o\bullet}(t)] + \\
    &Q(t)\mathbb{E}[\phi^\bullet(t)] \leq B + VC_{max} - \epsilon \sum\nolimits_{\smash{s=1}}^{\smash{|\textit{S}|}} D_s(t) - \epsilon Q(t)
\end{align*}
Take the expectation and sum up over the time slot $t \in \{0,1,2,..,T-1\}$, since $L(\bold{\Theta}(0)) = 0, L(\bold{\Theta}(T)) \geq 0$, we can obtain
\begin{align*}
    &0 \leq B'T - \epsilon\sum\nolimits_{t=0}^{T-1}\sum\nolimits_{\smash{s=1}}^{\smash{|\textit{S}|}} \mathbb{E}[D_s(t)] - \epsilon\sum\nolimits_{t=0}^{T-1}\mathbb{E}[Q(t)]
\end{align*}
where $B' = B + V(C_{max}-C_{min})$, then divide both sides by $T\epsilon$ and take $T \to \infty$:
\begin{equation*}
    \lim_{T\to\infty}\frac{1}{T}\left\{\sum\nolimits_{t=0}^{T-1}\sum\nolimits_{\smash{s=1}}^{\smash{|\textit{S}|}} \mathbb{E}[D_s(t)]+\sum\nolimits_{t=0}^{T-1}\mathbb{E}[Q(t)] \right\} \leq \frac{B'}{\epsilon}
\end{equation*}

\newpage
\bibliographystyle{IEEEtran}

\end{document}